\documentclass[10pt, conference, letterpaper]{IEEEtran}
\IEEEoverridecommandlockouts

\usepackage{multirow}
\usepackage{epsfig,amsmath,amssymb,latexsym}
\usepackage{graphicx,epsfig,color,epsf,psfrag,hhline,amsmath,amssymb,textcomp}

\usepackage[ruled]{algorithm2e}

\usepackage{float}
\usepackage{xcolor}
\usepackage{makecell}
\usepackage{comment}
\usepackage{tabularx} %extra features for tabular environment
\usepackage{mathtools}
\usepackage{cite} %takes care of citations
\usepackage[final]{hyperref} %adds hyper links inside the generated pdf file
\usepackage{amsthm}
\usepackage[utf8]{inputenc}
\usepackage[english]{babel}
\usepackage[export]{adjustbox}
\usepackage{subfig}

\hypersetup{
	colorlinks=true,       %false: boxed links; true: colored links
	linkcolor=blue,        %color of internal links
	citecolor=blue,        %color of links to bibliography
	filecolor=magenta,     %color of file links
	urlcolor=blue
}

\usepackage{epsfig,amsmath,amssymb,latexsym}
\usepackage{color,epsf,psfrag,hhline,textcomp}
\usepackage{graphics}

\usepackage{caption}
\usepackage{graphicx}
\usepackage{algorithmic}

\usepackage[ruled]{algorithm2e}

\def \_ {\rule{.3em}{.15ex}} %Get underscore by typing \_.

 \normalmarginpar
\newcommand {\mymarginpar}[1]{\marginpar{#1}}
\renewcommand {\marginpar}[1]{}

\def\_{\rule{.3em}{.15ex}}      % Get underscore by typing \_.

\newcommand{\ls}[1]
{\dimen0=\fontdimen6\the\font
	\lineskip=#1\dimen0
	\advance\lineskip.5\fontdimen5\the\font
	\advance\lineskip-\dimen0
	\lineskiplimit=.9\lineskip
	\baselineskip=\lineskip
	\advance\baselineskip\dimen0
	\normallineskip\lineskip
	\normallineskiplimit\lineskiplimit
	\normalbaselineskip\baselineskip
	\ignorespaces
}

\newcommand{\bearn}{\begin{eqnarray*}}
\newcommand{\eearn}{\end{eqnarray*}}
\newcommand{\barr}{\begin{array}}
\newcommand{\earr}{\end{array}}

\newcommand{\N}{{\cal N}}

\newtheorem{definition}{Definition}
\newtheorem{assumption}[definition]{Assumption}
\newtheorem{property}[definition]{Property}
\newtheorem{proposition}[definition]{Proposition}
\newtheorem{lemma}[definition]{Lemma}
\newtheorem{theorem}[definition]{Theorem}
\newtheorem{corollary}[definition]{Corollary}
\newtheorem{example}[definition]{Example}
\newtheorem{remark}[definition]{Remark}

\newcommand{\benum}{\begin{enumerate}}
\newcommand{\eenum}{\end{enumerate}}

\newcommand{\bdesc}{\begin{description}}
\newcommand{\edesc}{\end{description}}

\newcommand {\bfig}[2] {\begin{figure}
		\centering
		\includegraphics[width=#2]{#1}}
	\newcommand {\brotatefig}[2] {\begin{figure}[htbp]
			\centerline {
				\epsfig{figure={#1},clip=,angle=-90,width={#2}}}}
		\newcommand {\bfigfirst}[2] {\begin{figure}[h]
				\centerline {
					\setlength{\epsfxsize}{#2}
					\epsffile{#1}}}
			\newcommand {\efig}[2]{ \caption{#2}
				\label{fig:#1}
			\end{figure}
			\mymarginpar{fig:#1}}
		\newcommand {\erotatefig}[2]{ \caption{#2}
			\label{fig:#1}
		\end{figure}
		\mymarginpar{fig:#1}}

\newcommand{\btab}[1]{\begin{table}\centering\begin{tabular}{#1}}
\newcommand{\etab}[3]{\end{tabular}\caption[#3]{#2}\label{tab:#1}\end{table}\mymarginpar{tab:#1}\vspace{.1in}}

\newcommand{\btabular}[1]{\begin{center}\begin{tabular}{#1}}
\newcommand{\etabular}{\end{tabular}\end{center}}
\newcommand{\bdefin}[1]{\begin{definition}\mymarginpar{def:#1}\label{def:#1}}
\newcommand{\edefin}{\end{definition}}
\newcommand{\rdef}[1]{Definition \ref{def:#1}}
\newcommand{\bpro}[1]{\begin{property}\mymarginpar{pro:#1}\label{pro:#1}}
\newcommand{\epro}{\end{property}}

\newcommand{\bprop}[1]{\begin{proposition}\mymarginpar{prop:#1}\label{prop:#1}}
\newcommand{\eprop}{\end{proposition}}

\newcommand{\blem}[1]{\begin{lemma}\mymarginpar{lem:#1}\label{lem:#1}}
\newcommand{\elem}{\end{lemma}}
\newcommand{\rlem}[1]{Lemma \ref{lem:#1}}
\newcommand{\bass}[1]{\begin{assumption}\mymarginpar{the:#1}\label{ass:#1}}
\newcommand{\eass}{\end{assumption}}

\newcommand{\bthe}[1]{\begin{theorem}\mymarginpar{the:#1}\label{the:#1}}
\newcommand{\ethe}{\end{theorem}}
\newcommand{\rthe}[1]{Theorem \ref{the:#1}}
\newcommand{\bcor}[1]{\begin{corollary}\mymarginpar{cor:#1}\label{cor:#1}}
\newcommand{\ecor}{\end{corollary}}
\newcommand{\rcor}[1]{Corollary \ref{cor:#1}}
\newcommand{\bax}[1]{\begin{axiom}\mymarginpar{ax:#1}\label{ax:#1}}
\newcommand{\eax}{\vspace{-.1in} \end{axiom}}

\newcommand{\bex}[2]{\vspace{.1in}\begin{example}\mymarginpar{ex:#1}{\bf #2}\label{ex:#1} }
\newcommand{\eex}{\end{example}\vspace{.3cm}}

\newcommand{\brem}[1]{\begin{remark}\mymarginpar{rem:#1}\label{rem:#1}\em}
\newcommand{\erem}{\end{remark}}

\newcommand{\beq}[1]{\mymarginpar{eq:#1}\begin{equation}\label{eq:#1}}
\newcommand{\beqno}[1]{\mymarginpar{eq:#1}\begin{eqnarray}\nonumber}
\newcommand{\eeq}{\end{equation}}
\newcommand{\eeqno}{&&\end{eqnarray}}
\newcommand{\req}[1]{(\ref{eq:#1})}

\newcommand{\bear}[1]{\mymarginpar{eq:#1}\begin{eqnarray}\label{eq:#1}}
\newcommand{\bearno}[1]{\mymarginpar{eq:#1}\begin{eqnarray}\nonumber}
\newcommand{\eear}{\end{eqnarray}}
\newcommand{\eearno}{\end{eqnarray}}
\usepackage{environ}
\NewEnviron{es}{%
	\begin{equation}\begin{split}
	\BODY
	\end{split}\end{equation}
}
\newcommand{\bsel}{\left\{\begin{array}{cl}}
\newcommand{\esel}{\end{array}\right.}
\newcommand{\bmat}[1]{\left[\begin{array}{#1}}
\newcommand{\emat}{\end{array}\right]}
\newcommand{\bsec}[2]{\mymarginpar{sec:#2}\section{#1}\label{sec:#2}}

\newcommand{\bsubsec}[2]{\mymarginpar{subsec:#2}\subsection{#1}\label{subsec:#2}}

\newcommand{\bapp}{\begin{appendices}}
\newcommand{\eapp}{\end{appendices}}
\def\R{I\kern-0.30em R}
\def\N{I\kern-0.30em N}
\def\P{I\kern-0.30em P}
\def \pr{{\bf\sf P}}

\newcommand\sgo{(G(V,E), p_{U,W}(\cdot,\cdot))}
\newcommand \qq{q}
\newcommand \Q {{\bf Q}}

\begin{document}
\title{Generalized Modularity Embedding: a General Framework for Network Embedding}
\author{Cheng-Shang~Chang,~\IEEEmembership{Fellow,~IEEE,}
	Ching-Chu~Huang,
	Chia-Tai~Chang,\\
	Duan-Shin~Lee,~\IEEEmembership{Senior~Member,~IEEE,}
	and~Ping-En~Lu,~\IEEEmembership{Graduate~Student~Member,~IEEE}\\% <-this % stops a space
	Institute of Communications Engineering, National Tsing Hua University\\
	Hsinchu 30013, Taiwan, R.O.C.\\
	Email: cschang@ee.nthu.edu.tw; pri0106chu@gmail.com; s104064540@m104.nthu.edu.tw;\\
	lds@cs.nthu.edu.tw; j94223@gmail.com.
	\thanks{C.-S. Chang, C.-C. Huang, C.-T. Chang, D.-S. Lee and P.-E Lu are with the Institute of Communications Engineering, National Tsing Hua University, Hsinchu 30013, Taiwan, R.O.C. Email: cschang@ee.nthu.edu.tw; pri0106chu@gmail.com; s104064540@m104.nthu.edu.tw; lds@cs.nthu.edu.tw; j94223@gmail.com.}% <-this % stops a space
}

\date{\today}

%\markboth{Draft: May 11, 2001}
%{Murray and Balemi: Using the style file IEEEtran.sty} %!PN
%{Murray and Balemi: Using the Document Class IEEEtran.cls} %!PN
	
\maketitle

\begin{abstract}
The network embedding problem aims to map nodes that are similar to each other to vectors in a Euclidean space that are close to each other. Like centrality analysis (ranking) and community detection, network embedding is in general considered as an ill-posed problem, and its solution may depend on a person's view on this problem. In this paper, we adopt the framework of sampled graphs that treat a person's view as a sampling method for a network. The modularity for a sampled graph, called the generalized modularity in the paper, is a similarity matrix that has a specific probabilistic interpretation. One of the main contributions of this paper is to propose using the generalized modularity matrix for network embedding and show that the network embedding problem can be treated as a trace maximization problem like the community detection problem. Our generalized modularity embedding approach is very general and flexible. In particular, we show that the Laplacian eigenmaps is a special case of our generalized modularity embedding approach. Also, we show that dimensionality reduction can be done by using a particular sampled graph. Various experiments are conducted on real datasets to illustrate the effectiveness of our approach.
\end{abstract}


\bsec{Introduction}{labelofintroduction}

Network analysis has been a hot research topic as it has many important real-world applications such as web page ranking and recommender systems. There are several fundamental problems of network analysis (see, e.g., the book \cite{Newman2010}), including centrality analysis (ranking) and community detection (clustering). Centrality analysis is to identify ``important'' nodes in a network and community detection is to cluster nodes that are ``similar'' to each other into the same community. These problems are in general regarded as ill-posed problems as people might have different views on ``importance'' and ``similarity'' of the nodes in a network.

On the other hand, inspired by the success of deep learning (see, e.g., the book \cite{goodfellow2016deep}), many recent network embedding techniques have been proposed in the literature to learn meaningful networking representations (see, e.g., \cite{perozzi2014deepwalk, tang2015line, cao2015grarep, grover2016node2vec, wang2016structural, ou2016asymmetric, goyal2017graph, wang2017community, lai2017preserving}). The objective of network embedding is to map nodes that are ``similar'' to each other to vectors in a Euclidean space that are close to each other. By doing so, the intrinsic information of a network can be preserved. Like the node ranking (centrality) problem and the community detection problem in network analysis, the network embedding problem is also in general considered as an ill-posed problem and its solution may depend on a person's view on this problem and the task he/she would like to complete after network embedding. Most of the embedding methods aim to preserve $k$-order proximity, mainly focusing on the microscopic structure of the network. Early works like IsoMap \cite{tenenbaum2000global} and Laplacian eigenmaps \cite{belkin2002laplacian} were designed to preserve the {\em first}-order proximity between nodes. The {\em second}-order proximity was considered in \cite{tang2015line} and \cite{wang2016structural}. Furthermore, the works in \cite{perozzi2014deepwalk, ou2016asymmetric, cao2015grarep} extended the previous works to preserve even higher $k$-order proximity. In addition to these, some previous works also formulated the network embedding problem to preserve the community structure of a network \cite{wang2017community, li2013modularity}. More recently, the work \cite{lai2017preserving} aimed to preserve the global node ranking and local proximity of nodes.

Motivated by all the recent works of network analysis and network embedding, the objective of this paper is to provide a general framework for the network embedding problem so that the ill-posed problem can be treated in a formal and unified manner with probabilistic insights. Our approach for this is to use the probabilistic framework in \cite{chang2013relative, chang2017probabilistic}. In that framework, a person's view might be represented by the sampling method he/she uses for sampling a network, and each sampling method renders a sampled graph that is characterized by a bivariate distribution $p(u,w)$ representing the probability that the two nodes $u$ and $w$ appear together in a sample. The marginal distribution of the bivariate distribution $p(u,w)$, denoted by $P_U(u)$, is then the probability that node $u$ is sampled and that in turn can be used for representing the ``importance'' of the node. The {\em covariance} between two nodes $u$ and $w$, denoted by $q(u,w)$, is
$$q(u,w)=p(u,w)-p_U(u)p_W(w).$$
Intuitively, two nodes that are positively (negatively) correlated are ``similar'' (resp. ``dissimilar''). The matrix $\Q=(q(u,w))$ is called the {\em generalized modularity} matrix as it reduces to the Newman's modularity matrix in \cite{Newman04} when edge sampling is used for generating the sampled graph. The stability matrix in \cite{Lambiotte2010, Delvenne2010} that was used for addressing the problem of resolution limit \cite{FB07} is also a special case of the generalized modularity matrix when one uses the continuous-time random walk for sampling \cite{chang2017probabilistic}. Community detection in a sampled graph is then formulated as a modularity maximization problem that finds a $K$-way partition $S_1, \ldots, S_K$ of nodes to maximizes the modularity $\sum_{k=1}^K q(S_k,S_k)$. Such a modularity maximization problem is equivalent to the trace maximization problem that maximizes $\mbox{tr}(H^T \Q H)$ among the class of partition matrix $H$.

In view of the probabilistic framework in \cite{chang2013relative, chang2017probabilistic}, we propose using the generalized modularity matrix $\Q$ as the ``similarity'' matrix in a network embedding problem. By doing so, we show that {\em the network embedding problem can be treated as a trace maximization problem like the community detection problem}. As such, it is possible to solve the network embedding problem by using the well-known community detection algorithms in the literature, e.g., the partitional algorithm \cite{chang2017probabilistic} and the fast unfolding algorithm \cite{blondel2008fast}. On the other hand, as each sampling method leads to a sampled graph, our generalized modularity embedding approach is very general and flexible. In particular, we show that the Laplacian eigenmaps is a special case of our generalized modularity embedding for a particular sampled graph. Also, we show that dimensionality reduction (and PCA) can be done by using a particular sampled graph. Based on the framework, network embedding algorithms can be developed and written in codes by using sampled graphs as inputs. If one is not satisfied with the network embedding result from a sampled graph, one can simply choose another sampling method for the graph and try out the analysis again. There is no need to rewrite the codes for the network embedding algorithm.

The rest of this paper is organized as follows. In Section \ref{sec:framework}, we introduce sampled graphs and the generalized modularity, and formulate the community detection problem and the network embedding problem. In Section \ref{sec:extensions}, we extend generalized modularity embedding for data points in a semi-metric space. We show the connections to the Laplacian eigenmaps and dimensionality reduction. In Section \ref{sec:softmax}, we consider the nonnegative embedding problem and provide the \textit{softmax embedding/clustering algorithm} to solve the nonnegative embedding problem and the community detection problem simultaneously. In Section \ref{sec:experiments}, we report the experimental results by comparing the performance of our work with several baseline embedding approaches. The paper is concluded in Section \ref{sec:conclusion}.

\bsec{The probabilistic framework for generalized modularity embedding}{framework}

\bsubsec{The generalized modularity matrix of a sampled graph}{resampled}

In \cite{chang2013relative, chang2017probabilistic}, a probabilistic framework for structural analysis in undirected/directed networks was proposed. The main idea in that framework is to sample a network by randomly selecting a path in the network. A network with a path sampling distribution is then called a {\em sampled graph} in \cite{chang2013relative, chang2017probabilistic} that can, in turn, be used for structural analysis of the network, including centrality and community. Specifically, suppose a network is modeled by a graph $G(V,E)$, where $V$ denotes the set of vertices (nodes) in the graph and $E$ denotes the set of edges (links) in the graph. Let $n=|V|$ be the number of vertices in the graph and index the $n$ vertices from $1,2,\ldots, n$. Let $R_{u,w}$ be the set of (directed) paths from $u$ to $w$ and $R=\cup_{u,w \in V} R_{u,w}$ be the set of paths in the graph $G(V,E)$. According to a probability mass function $p(\cdot)$, called the {\em path sampling distribution}, a path $r \in R$ is selected at random with probability $p(r)$ (see, e.g., \cite{chang2013relative, chang2017probabilistic} for examples of sampling methods).

Let $U$ (resp. $W$) be the starting (resp. ending) node of a randomly selected path by using the path sampling distribution $p(\cdot)$. Then the bivariate distribution
\beq{frame1111}
p_{U,W}(u,w)=\pr(U=u,W=w)=\sum_{r\in R_{u,w}} p(r)
\eeq
is the probability that the ordered pair of two nodes $(u,w)$ is selected.

\bdefin{sampled}{\bf (Sampled graph \cite{chang2013relative, chang2017probabilistic})} A graph $G(V,E)$ that is sampled by randomly selecting an ordered pair of two nodes $(U,W)$ according to a specific bivariate distribution $p_{U,W}(\cdot,\cdot)$ is called a {\em sampled graph} and it is denoted by the two-tuple $\sgo$.
\edefin

Let $p_U(u)$ (resp. $p_W(w)$) be the marginal distribution of the random variable $U$ (resp. $W$), i.e.,
\beq{frame2222}
p_U(u)=\pr(U=u)= \sum_{w=1}^n p_{U,W}(u,w),
\eeq
and
\beq{frame3333}
p_W(w)= \pr(W=w)=\sum_{u=1}^n p_{U,W}(u,w).
\eeq
Then $p_U(u)$ is the probability that node $u$ is selected as a starting node of a path and it can be viewed as an out-centrality of $u$. On the other hand, $p_W(w)$ is the probability that node $w$ is selected as an ending node of a path and it can be viewed as an in-centrality of $w$. The in-centrality and the out-centrality are in general not the same. Clearly, if the bivariate distribution $p_{U,W}(\cdot,\cdot)$ is symmetric, then the in-centrality and the out-centrality are the same. A recent advance in \cite{chang2017probabilistic} shows that one does not need a symmetric bivariate distribution to ensure the equality between the in-centrality and the out-centrality. In particular, for the Markov chain sampling methods, one still has $p_U(u)=p_W(u)$ and the in-centrality and the out-centrality are the same. In that case, we will simply refer $P_U(u)$ as the centrality of node $u$.
\begin{definition}\label{def:covariance}{\bf (Covariance, Community, and Modularity \cite{chang2013relative, chang2017probabilistic}))}
	For a sampled graph $\sgo$, the covariance between two nodes $u$ and $w$ is defined as follows:
	\begin{equation}\label{eq:exp8888}
		\qq (u,w)=p_{U,W} (u,w)-p_{U}(u) p_{W}(w).
	\end{equation}
	Define the generalized modularity matrix $\Q = (\qq (u,w))$ be the $n\times n$ matrix with its $(u,w)^{th}$ element being $\qq(u,w)$.
	%where $p_{\lambda}(x,y)$ is in (\ref{eq:exp5555}).
	Moreover, the covariance between two sets $S_1$ and $S_2$ is defined as follows:
	\begin{equation}\label{eq:exp9999}
		\qq (S_1,S_2)=\sum_{u \in S_1}\sum_{w \in S_2}\qq (u,w).
	\end{equation}
	Two sets $S_1$ and $S_2$ are said to be positively correlated if $\qq (S_1,S_2)\ge 0$.
	In particular, if a subset of nodes $S \subset V$ is positively correlated to itself, i.e., $\qq(S,S)\ge 0$, then it is
	called
	a {\em community} or a {\em cluster}.
	Let ${\cal P}=\{S_k,k=1,2, \ldots, K\}$, be a partition of $V$, i.e.,
	$S_k \cap S_{k^\prime}$ is an empty set for $k \ne k^\prime$ and $\cup_{k=1}^K S_k=V$.
	The modularity $Q({\cal P})$ with respect to the partition $S_k$, $k=1,2, \ldots, K$, is defined as
	\begin{equation}\label{eq:index1111}
		Q({\cal P})=\sum_{k=1}^K \qq(S_k,S_k).
	\end{equation}
\end{definition}
As pointed out in \cite{chang2013relative, chang2017probabilistic}, the original modularity defined in \cite{Newman04} is a special case of the (generalized) modularity in \rdef{covariance} when the uniform edge sampling is used to generate the sampled graph.

\bsubsec{Community detection}{detection}

There are many physical interpretations (and equivalent statements) for the definition of a community in \cite{chang2013relative, chang2017probabilistic}. Moreover, as pointed out in the book \cite{Newman2010}, the physical meaning of the modularity with respect to a partition of a graph is how much it differs from that partition of a random graph generated by the configuration model. As such, a good partition of a graph should have large modularity. In view of this, one can then tackle the community detection/clustering problem by looking for algorithms that yield large modularity.

It is well-known that the modularity maximization problem in \req{index1111} is NP-complete and is related to the trace maximization problem \cite{shi2000normalized, yu2003multiclass, Dhillon04b}. One can view an $n \times K$ partition matrix $H$ as a bi-adjacency matrix of a bipartite graph that maps the $n$ nodes to the $K$ sets of a $K$-way partition $\{S_1, S_2, \ldots ,S_K\}$, and thus there is a one-to-one mapping between a partition matrix and a $K$-way partition. Let $H_k$ be the $k^{th}$ column vector of $H$ (that represents the set of data points in $S_k$). To maximize the modularity in \req{index1111}, it is thus equivalent to consider the following optimization problem:
\bear{opti1111h}
&\max&\sum_{k=1}^K H^T_k \Q H_k \\
&s.t. &H \in \{0,1\}^{n \times K}, \nonumber\\
&& {\bf 1}_n^T H>0,\nonumber \\
&& H {\bf 1}_K ={\bf 1}_n, \nonumber
\eear
where ${\bf 1}_K$ and ${\bf 1}_n$ are $K$-dimensional and $n$-dimensional column vectors with
all its elements being 1. Such an optimization problem can also be written as a trace maximization problem as follows:
\bear{opti1111trall}
&\max &\mbox{tr}(H^T \Q H) \\
&s.t. &H \in \{0,1\}^{n \times K}, \nonumber\\
&& {\bf 1}_n^T H>0,\nonumber \\
&& H {\bf 1}_K ={\bf 1}_n. \nonumber
\eear

Instead of maximizing modularity, another alternative for community detection is to maximize the {\em normalized} modularity \cite{chang2016mathematical} as follows:
\bear{opti1111trnor}
&\max&\sum_{k=1}^K \frac{H^T_k \Q H_k}{ H^T_k H_k}\\
&s.t. &H \in \{0,1\}^{n \times K}, \nonumber\\
&& {\bf 1}_n^T H>0,\nonumber \\
&& H {\bf 1}_K ={\bf 1}_n. \nonumber
\eear
As shown in \cite{yu2003multiclass}, one can use the scaled partition matrix
\beq{opti2222h}
\tilde H=H(H^T H)^{-\frac{1}{2}}
\eeq
to represent the objective function in \req{opti1111trnor} by
$\mbox{tr}(\tilde H^T \Q \tilde H)$.
Since
$$\tilde H^T \tilde H =(H^T H)^{-\frac{1}{2}} H^T H (H^T H)^{-\frac{1}{2}}={\bf I}_K,$$
one can relax the integer constraints in the optimization problem in \req{opti1111trnor} and consider the following trace maximization problem:
\bear{opti1111r}
&&\max \mbox{tr}(\tilde H^T \Q \tilde H) \\
&&s.t. \quad \tilde H^T \tilde H ={\bf I}_K. \nonumber
\eear

Since the modularity matrix $\Q$ is a real symmetric matrix for an undirected network, it is diagonalizable. As stated in \cite{von2007tutorial}, a version of the Rayleigh-Ritz theorem shows that the solution of the trace maximization problem in \req{opti1111r} is given by choosing $\tilde H$ as the matrix which contains the largest $K$ eigenvectors of $\Q$ as columns. As the trace maximization problem in \req{opti1111r} is a relaxation of the optimization problem in \req{opti1111trnor}, this also gives an upper bound on the objective value of the optimization problem in \req{opti1111trnor}. Specifically, let $\lambda_1, \lambda_2, \ldots, \lambda_K$ be the $K$ largest eigenvalues of the matrix $\Q$. Then the objective value of the optimization problem in \req{opti1111trnor} is bounded above by $\sum_{k=1}^K \lambda_k$.

The rest of the problem is to map the solution for the trace optimization problem in \req{opti1111r} back to the solution space for the optimization problem in \req{opti1111trnor} (that requires the solutions to be binary).

\bsubsec{Modularity embedding}{modularity embedding}

The network embedding problem is quite similar to the community detection (clustering) problem. The community detection problem is to assign nodes that are similar to each other in the same cluster and nodes that are dissimilar to each other in different clusters. On the other hand, the network embedding problem is to map nodes that are similar to each other to vectors in a Euclidean space that are close to each other. In the previous section, we use the (generalized) modularity as the similarity measure for community detection in a sampled graph. Here we also use the (generalized) modularity as the similarity measure for network embedding in a sampled graph. We will show that the network embedding problem can also be formulated as a modularity maximization problem. Specifically, let $h_u=(h_{u,1}, h_{u,2}, \ldots, h_{u,K})^T$ be the vector mapped by node $u$ in $\mathbb{R}^K$, and $||h_u-h_w||^2$ be the squared Euclidean distance between $u$ and $w$. In order to map nodes that are similar to each other to vectors that are close to each other, we consider the problem that minimizes the following weighted distance:
\beq{embed1111}
\sum_{u=1}^n \sum_{w=1}^n \qq(u,w) ||h_u-h_w||^2.
\eeq
Such a network embedding approach was previously used in \cite{li2013modularity} when $\qq(u,w)$ is the original modularity defined in \cite{Newman04}. To understand the intuition of the minimization problem in \req{embed1111}, note that $-1 \le \qq(u,w) \le 1$. Two nodes with a positive (resp. negative) covariance should be mapped to two points with a small (resp. large) distance. The embedding vector $h_u=(h_{u,1}, h_{u,2}, \ldots, h_{u,K})^T$ can be viewed as the ``feature'' vector of node $u$ and $h_{u,k}$ is its $k^{th}$ feature. In practice, it is preferable to have uncorrelated features. For this, we add the constraints
\beq{embed1122}
\sum_{u=1}^n h_{u,k_1} h_{u,k_2} =0,
\eeq
for all $k_1 \ne k_2$. Also, to have bounded values for these features, we also add the constraints
\beq{embed1133}
\sum_{u=1}^n h_{u,k} h_{u,k} =1,
\eeq
for all $k$. Thus, our generalized modularity embedding problem is to solve the minimization problem in \req{embed1111} under the constraints in \req{embed1122} and \req{embed1133}.

In the following theorem, we show two equivalent statements for the generalized modularity embedding problem.
Its proof can be found in Appendix A.
\bthe{embed}
Let $\Q =(\qq_{u,w})$ be an $n \times n$ symmetric matrix with all its row sums and column sums being 0, and
$H$ be the $n \times K$ matrix with its $u^{th}$ row being $h_u$.
\begin{description}
	\item[(i)] The generalized modularity embedding problem in \req{embed1111} with the constraints in \req{embed1122} and \req{embed1133} is equivalent to the following optimization problem:
	\bear{opti1111tr}
	&\max &\mbox{tr}(H^T \Q H)
	\\
	&s.t. &H^T H={\bf I}_K,
	\eear
	where ${\bf I}_K$ is the $K \times K$ identify matrix.
	\item[(ii)] The generalized modularity embedding problem in \req{embed1111} with the constraints in \req{embed1122} and \req{embed1133} is equivalent to the following optimization problem:
	\bear{opti1111decom}
	&\min &||\Q- H H^T||_2^2\\
	&s.t. &H^T H={\bf I}_K,
	\eear
	where $||A||_2$ is the Frobenius norm of the matrix $A$.
\end{description}
\ethe

From \rthe{embed}(i), we know that solving the generalized modularity embedding problem is equivalent to solving the trace maximization problem in \req{opti1111tr}. Since the optimization problem in \req{opti1111tr} is the same as the relaxed version of the normalized modularity maximization problem in \req{opti1111r}, we have the following corollary.
\bcor{embedcor}
For the generalized modularity embedding problem, let $\lambda_1, \lambda_2, \ldots, \lambda_K$ be the $K$ largest eigenvalues of the matrix $\Q$ and $v_k=(v_{k, 1}, v_{k, 2}, \ldots, v_{k, n})^T$ be the eigenvector of $\Q$ corresponding to the eigenvalue $\lambda_k$. Then $h_u=( v_{1,u}, v_{2,u}, \ldots, v_{K,u})^T$, $u=1,2, \ldots, n$, are the optimal embedding vectors.
\ecor

\bsec{Extensions and applications of generalized modularity embedding}{extensions}

\bsubsec{Modularity embedding for data points in a semi-metric space}{semi}

In this section, we extend our generalized modularity embedding method for embedding data points in a semi-metric space. For this, we consider a set of $n$ data points associated with a distance measure $d(u,w)$. The distance measure $d(\cdot, \cdot)$ is assumed to be a {\em semi-metric} and it satisfies the following three properties:
\begin{description}
	\item[(D1)] (Nonnegativity) $d(u,w) \ge 0$.
	\item[(D2)] (Null condition) $d(u,u)=0$.
	\item[(D3)] (Symmetry) $d(u,w)=d(w,u)$.
\end{description}

The semi-metric assumption is weaker than the metric assumption in \cite{chang2016mathematical}, where the distance measure is assumed to satisfy the triangular inequality.

Given a semi-metric $d(\cdot,\cdot)$, it was shown in \cite{chang2017k} that there is an induced semi-cohesion measure as follows:
\begin{eqnarray}
	\gamma(u,w)=\frac{1}{n}\sum_{u_2 \in \Omega} d(u_2,w)+\frac{1}{n} \sum_{u_1 \in \Omega}d(u,u_1)\nonumber\\
	-\frac{1}{n^2} \sum_{u_2 \in \Omega}\sum_{u_1 \in \Omega}d(u_2,u_1)-d(u,w).
	\label{eq:csim7777}
\end{eqnarray}
Moreover, the induced semi-cohesion measure satisfies the following three properties:
\begin{description}
	\item[(C1)] (Symmetry) $\gamma(u,w)=\gamma(w,u)$.
	\item[(C2)] (Null condition) $\sum_{w=1}^n\gamma (u,w)=0$.
	\item[(C3)] (Nonnegativity)
	\begin{equation}
		\label{eq:cmeas1111}
		\gamma(u,u)+\gamma (w,w) \ge 2\gamma (u, w).
	\end{equation}
\end{description}
On the other hand, for any semi-cohesion measure that satisfies (C1)--(C3), there is also an induced semi-metric as follows:
\begin{equation}
	\label{eq:cind1111}
	d(u,w)=(\gamma(u,u)+\gamma(w,w))/2 -\gamma(u,w).
\end{equation}
Thus, there is a one-to-one mapping (a duality result) between a semi-metric and a semi-cohesion measure. In this paper, we will simply say data points are in a semi-metric space if there is either a semi-cohesion measure or a semi-metric associated with these data points.

We note that a semi-cohesion measure (matrix) that satisfies (C1)--(C3) may not be positive semi-definite (see, e.g., Example 20 in \cite{chang2016mathematical}).

The set of $n$ nodes with the semi-metric $d(u,w)$ can be viewed as a complete graph with $n$ nodes, where the edge weight between nodes $u$ and $w$ is $d(u,w)$. For such a complete graph, the following bivariate distribution was proposed in \cite{chang2018e}:
\beq{exp1111c}
p_{U,W}(u,w)=\frac{\exp(\theta \cdot d(u,w))}{\sum_{u_1}\sum_{u_2}\exp(\theta \cdot d(u_1,u_2))}.
\eeq

For the bivariate distribution in \req{exp1111c}, let us consider the covariance measure $\qq(u,w)$ in \req{exp8888} when $\theta$ is very small. Using the first order approximation $e^{\theta z}\approx 1+\theta z + o(\theta)$ in \req{exp8888} yields
\begin{eqnarray*}
	\qq(u,w) &\approx&
	(-\theta) \Big (\frac{1}{n}\sum_{u_2 \in V} d(u_2,w)+\frac{1}{n} \sum_{u_1 \in V}d(u,u_1)\nonumber\\
	&&
	-\frac{1}{n^2} \sum_{u_2 \in V}\sum_{u_1 \in V}d(u_2,u_1)-d(u,w) \Big) +o(\theta).
	\label{eq:expcoh2222}
\end{eqnarray*}
Thus, when we choose a very small negative $\theta$, the covariance measure $\qq(u,w)$ is proportional to the semi-cohesion measure in \req{csim7777}.

Since the covariance measure $\qq(u,w)$ is proportional to the cohesion measure $\gamma(u,w)$ for a small negative $\theta$, the trace maximization problem in \req{opti1111tr} is is equivalent to the following trace maximization problem:
\bear{opti1111trg}
&\max &\mbox{tr}(H^T \Gamma H),\\
&s.t. &H^T H={\bf I}_K,
\eear
where $\Gamma$ is the $n \times n$ matrix with its $(u,w)^{th}$ element being $\gamma(u,w)$. In view of \rcor{embedcor}, embedding data points in a semi-metric space to $\mathbb{R}^{K}$ can be solved by finding the $K$ largest eigenvalues of the matrix $\Gamma$ and the corresponding eigenvectors.

\bsubsec{The Laplacian eigenmaps as a special case of generalized modularity embedding}{ne}

The Laplacian eigenmaps in \cite{belkin2002laplacian} that uses the eigenvectors of the Laplacian matrix of a graph for network embedding has been widely used in the literature. In this section, we show that the Laplacian eigenmaps for network embedding in \cite{belkin2002laplacian} is, in fact, a special case of the generalized modularity embedding by using the effective resistance distance of a graph.

The (graph) Laplacian of an undirected network $G=(V,E)$, denoted by $L$, is defined as $D-A$, where $D=diag(w_1, w_2, \ldots, w_n)$ is the $n \times n$ diagonal matrix with the diagonal elements being the degrees of the $n$ nodes. It is well-known that the Laplacian $L$ is symmetric and positive semi-definite (see, e.g., \cite{spielman2007spectral, Newman2010}). As such, all its $n$ eigenvalues, $\beta_i, i=1,2, \ldots, n$, are nonnegative. Without loss of generality, we order the $n$ eigenvalues such that
$$\beta_1 \le \beta_2 \le \ldots \le \beta_n.$$
Moreover, there exists an orthonormal basis of $n \times 1$ column vectors $\{{\bf Z}_1,{\bf Z}_2, \ldots, {\bf Z}_n\}$ such that ${\bf Z}_i$ is the eigenvector of $L$ corresponding to the eigenvalue $\beta_i$, i.e.,
$$L {\bf Z}_i=\beta_i {\bf Z}_i.$$
As such, the Laplacian has the following eigendecomposition:
\beq{spectral1111}
L=\sum_{i=1}^n \beta_i {\bf Z}_i {\bf Z}_i^T,
\eeq
where ${\bf Z}_i^T$ is the transpose of ${\bf Z}_i$.

It is well-known (see, e.g., \cite{Newman2010}) that the smallest eigenvalue, $\beta_1$, is 0, and ${\bf Z}_1$ is the $n \times 1$ column vector with all its elements being $1/\sqrt{n}$. If, furthermore, the graph $G$ is connected, then the second smallest eigenvalue, $\beta_{2}$, known as the {\em algebraic connectivity}, is always positive. Thus, for a connected graph, one can define the pseudo-inverse of the Laplacian $L$ as follows:
\beq{spectral2222}
\Gamma= \sum_{i=2}^{n} \frac{1}{\beta_i} {\bf Z}_i {\bf Z}_i^T.
\eeq
The resistance distance between two nodes $u$ and $v$, denoted by $d(u,v)$, is then defined as
\beq{spectral3333}
d(u,v)=\gamma(u,u)+\gamma(v,v)-2 \gamma(u,v),
\eeq
where $\gamma(u,v)$ is the $(u,v)^{th}$ element of the matrix $\Gamma$. Using \req{spectral2222} yields
\beq{spectral4444}
d(u,v)=\sum_{i=2}^{n} \frac{(z_{i,u}-z_{i,v})^2}{\beta_i},
\eeq
where $z_{i,u}$ is the $u^{th}$ element of the vector ${\bf Z}_i$.
The resistance distance is known to be a metric that satisfies the triangular inequality. It then follows from the duality result that $\gamma(u,v)$ is a cohesion measure.

Now the $n$ nodes in the network $G=(V,E)$ can be viewed as $n$ data points in a semi-metric space with the
resistance distance as their distance measure. Using the embedding for data points in a semi-metric space to $\mathbb{R}^K$ in \req{opti1111trg},
we then have (from \rcor{embedcor}) that
\beq{feature1111}
h_u =(z_{2,u}, z_{3,u}, \ldots, z_{K+1,u}),
\eeq
$u=1,2, \ldots, n$, are the optimal embedding vectors for the generalized modularity embedding problem.

\bsubsec{Dimensionality reduction and connections to PCA}{PCA}

In this section, we show that the generalized modularity embedding can be used for dimensionality reduction and its connections to the principal component analysis (PCA). Consider $n$ data points in a Euclidean space $\mathbb{R}^{p}$. Let $d(u, w)$ be one-half of the squared Euclidean distance between any two points $u$ and $w$ i.e.,
\beq{PCA0011}
d(u, w)=\frac{1}{2}(u-w)^T(u-w).
\eeq
Then the $n$ data points in a Euclidean space $\mathbb{R}^{p}$ can be considered as $n$ data points in a semi-metric space with the semi-metric in \req{PCA0011}. It is straightforward to show that the semi-cohesion measure $\gamma(\cdot,\cdot)$ in \req{csim7777}
has the following form:
\begin{equation}
	\gamma(u, w ) = (u-c)^T(w-c),
\end{equation}
where
\begin{equation}
	c = \frac{1}{n}\sum_{u_2=1}^n u_2=\frac{1}{n}\sum_{u_1=1}^n u_1
\end{equation}
is the centroid of all the points in the dataset.

Without loss of generality, one can always subtract every point from its centroid to obtain another zero-mean dataset. For such a zero-mean dataset, the cohesion measure of two points is simply the inner product between the two points, i.e.,
\begin{equation}
	\gamma(u, w) = u^T w.
\end{equation}
Thus, the semi-cohesion matrix $\Gamma=(\gamma(u, w))$ obtained from using the {\em squared} Euclidean distance is a positive semi-definite matrix. As such, there are $n$ nonnegative eigenvalues. Let $\lambda_1\geq\lambda_2\geq\ldots\geq\lambda_n\geq0$ be the $n$ nonnegative eigenvalues of $\Gamma$, and $v_k$ be the corresponding (normalized) eigenvector for $\lambda_k$ with $\|v_k\|=1$. Thus, for $k=1,2, \ldots, n$,
$$\Gamma v_k=\lambda_k v_k.$$
Thus, the solution of the trace maximization problem in \req{opti1111trg} (cf. \rcor{embedcor}) leads to the following modularity embedding
\beq{pca3355}
h_u=(v_{1,u}, v_{2,u}, \ldots, v_{K,u})^T.
\eeq
This is almost the same (up to a scaling factor) as the following $K$-vector obtained by PCA for dimensionality reduction:
$$(\sqrt{\lambda_1} v_{1,u}, \sqrt{\lambda_2} v_{2,u}, \ldots, \sqrt{\lambda_K} v_{K,u})^T.$$
Thus, PCA is very closely related to the generalized modularity embedding when we use the sampled graph in \req{exp1111c} with a very small negative $\theta$.

\bsec{Nonnegative embedding and clustering}{softmax}

Both the embedding problem in \req{opti1111tr} and the community detection problem in \req{opti1111trall} are formulated as the trace maximization problem under certain constraints. In view of the similarity between these two problems, there is no doubt that these two problems are closely related. One interesting question is whether these two problems can be simultaneously solved. For this, let us consider the following relaxed version of the trace optimization problem in \req{opti1111trall}:
\bear{opti1111trplus}
&\max &\mbox{tr}(H^T \Q H) \\
&s.t. &H \in {\mathbb{R}^+}^{n \times K}, \nonumber\\
&& H {\bf 1}_K ={\bf 1}_n. \nonumber
\eear
There $H$ is only a nonnegative matrix with every row sum being 1. In other words, every row is a probability distribution, i.e.,
\bear{non1111}
&&h_{u,k} \ge 0, \quad \forall \;u, k, \label{eq:non1111a} \\
&&\sum_{u=1}^n h_{u,k} =1, \quad \forall \; k. \label{eq:non1111b}
\eear
If we use the trace optimization problem in \req{opti1111trplus} for embedding, then every node is mapped to a probability distribution. We will call such an embedding the nonnegative embedding problem. Once a good solution of the nonnegative embedding problem is obtained, one can then map that solution back to the solution space for the optimization problem in \req{opti1111trall} that requires the solutions to be binary. Intuitively, one can view $h_{u,k}$ as the probability that node $u$ belongs to cluster $k$. For the community detection problem, one can simply assign node $u$ to the cluster with the largest probability.

Now we propose an iterative algorithm for the nonnegative embedding problem, called the {\em softmax embedding/clustering algorithm} in Algorithm \ref{alg:Softmax}, based on the softmax function \cite{SoftmaxFunc}. The softmax function maps a $K$-dimensional vector of arbitrary real values to a $K$-dimensional probability vector. The algorithm starts from a non-uniform probability mass function for the assignment of each data point to the $K$ clusters. Specifically, let $h_{u,k}$ denote the probability that node $i$ is in cluster $k$. Then we repeatedly feed each point to the algorithm to learn the probabilities $h_{u,k}'s$. When point $i$ is presented to the algorithm, its expected covariance $z_{u,k}$ to cluster $k$ is computed for $k=1,2,\ldots, K$. Instead of assigning point $i$ to the cluster with the largest positive covariance (the simple maximum assignment in the literature), Algorithm 1 uses a softmax function to update $h_{u,k}'s$. Such a softmax update increases (resp. decreases) the confidence of the assignment of point $i$ to clusters with positive (resp. negative) covariances. The ``training'' process is repeated until the objective value $\mbox{tr}(H^T \Q H)$ converges to a local optimum. The algorithm then outputs the corresponding embedding vector for each data point.
\begin{algorithm}[ht]
	\KwIn{A symmetric modularity matrix $\Q=(q(u,w))$, the number of clusters $K$, the inverse temperature $\theta>0$.}
	\KwOut{A probabilistic embedding of data points $\{h_{u,k},u=1,2,\ldots,n,\;k=1,2,\ldots, K\}$.}
	
	\noindent {\bf (1)} Set $q(u,u)=0$ for all $u$.
	
	\noindent {\bf (2)} Initially, each node $u$ is assigned with a (non-uniform) probability mass function $h_{u,k}$, $k=1, 2, \ldots, K$ that denotes the probability for node $u$ to be in cluster $k$.
	
	\noindent {\bf (3)} For $u=1, 2, \ldots, n$
	
	\noindent {\bf (4)} For $k=1, 2, \ldots, K$
	
	\noindent {\bf (5)} Compute $z_{u,k}=\sum_{w\neq u}q(w,u) h_{w,k}$.
	
	\noindent {\bf (6)} Let $\tilde h_{u,k}=e^{\theta z_{u,k}}h_{u,k}$, and $c=\frac{1}{\sum_{\ell=1}^K \tilde h_{u,\ell}}$.
	
	\noindent {\bf (7)} Update $h_{u,k} \Leftarrow c \cdot {\tilde h_{u,k}}$.
	
	\noindent {\bf (8)} Repeat from Step 3 until there is no further change.
	
	\caption{The Softmax Embedding/Clustering Algorithm}
	\label{alg:Softmax}
\end{algorithm}

Now we show that Algorithm \ref{alg:Softmax} converges to a local maximum of the objective function $\mbox{tr}(H^T \Q H)$.
Its proof can be found in Appendix B.
\bthe{ObjIncreasing}
Given a symmetric matrix $\Q=(q(u,w))$ with $q(u,u)=0$ for all $u$, the following objective value
\beq{objc1111}
\mbox{tr}(H^T \Q H)= \sum\limits_{k=1}^K\sum\limits_{u=1}^n\sum\limits_{w=1}^n q(u,w)h_{u,k}h_{w,k}
\eeq
is increasing after each update in Algorithm \ref{alg:Softmax}. Thus, the objective values converge monotonically to a finite constant.
\ethe

\bsec{Experiments}{experiments}

In this section, we evaluate the performance of our \textit{Generalized Modularity Embedding (GME)} on the multi-classification problem. We consider three datasets: one synthetic dataset (\textit{Six Clusters} \cite{chang2017probabilistic}) and two real-world networks (\textit{Amazon}, and \textit{Flickr} \cite{flickrdataset}) from the Standford Network Analysis Project Collection (SNAP) \cite{yang2015defining}. As in \cite{chang2017probabilistic}, we use the \textit{maximum independent set (MIS)} algorithm to prune the two real networks so that they contain {\em disjoint} ground communities. An overview of these three networks is summarized in Table 1.\\
\begin{table}[ht]
	\centering
	Table 1. Dataset Statistics \label{tab:expdatasets}\\
	\begin{tabular}{|c||c|c|c|}
		\hline
		& \multicolumn{1}{c|}{\textit{synthetic}}& \multicolumn{2}{c|}{\textit{real-world}}\\ \hline
		Name & SIX CLUSTERS & AMAZON & FLICKR \\ \hline
		$|V|$ & 1,500 & 14,983 & 7575 \\ \hline
		$|E|$ & 49,431 & 44,039 & 242,146 \\ \hline
		Avg. degree & 65.91 & 5.88 & 63.93		\\ \hline
		\#Labels & 6 & 1,174 & 9 \\ \hline
	\end{tabular}
\end{table}
For each dataset, $1\%$ to $10\%$ or $10\%$ to $90\%$ of the labeled points are randomly chosen for training, and the rest of the data points are for evaluation.

For \textit{GME}, we first use a random walk on an undirected network (see, e.g., \cite{chang2013relative}) with path length 2, 3 or 4, to generate the generalized modularity matrix $\Q$. Then we solve the embedding problem by eigendecomposition as stated in \rcor{embedcor}. The embedding dimensionality $d$ is selected from the spectral gap of the (sorted) eigenvalues of the generalized modularity matrix $\Q$. For the three datasets, we have $d=6$ in \textit{Six Clusters}, $d=19$ in \textit{Flickr}, and $d=1200$ in \textit{Amazon}. We then recompose the covariance matrix $\Q^\prime$ from $HH^T$ and apply the softmax clustering algorithm in Algorithm \ref{alg:Softmax} to the classification problem with the input matrix $\Q^\prime$. The probability mass functions of the labeled training points are assigned with probability 1 to their corresponding labels and they remain unchanged during the iterations of Algorithm \ref{alg:Softmax}.

We compare \textit{GME} with the following four baseline algorithms: \textit{LINE} \cite{tang2015line}, \textit{DeepWalk} \cite{perozzi2014deepwalk}, \textit{node2vec} \cite{grover2016node2vec} and \textit{GraRep} \cite{cao2015grarep}. The dimensionality of the embedding vectors $d$ is set to be 128 for these four baseline algorithms. For \textit{Line}, we set a negative ratio $K=5$. For \textit{DeepWalk} and \textit{node2vec}, the number of random walks at each vertex is 10, the length of each random walk is 80, and the window size of skip-gram is 10. For $in$-$out$ $parameter$ $p$ and $return$ $parameter$ $q$ in \textit{node2vec}, we set $p=1$ and $q=1.5$ to capture structural equivalence since large $q$ approximates the BFS behavior. For $GraRep$, we set the k-step transition probability matrix with $k=5$.

To evaluate the performance of embedding tasks on node classification, we use the two F-measures: \textit{Micro-$F_1$} and \textit{Macro-$F_1$} defined below:
\bear{defmacro}
Macro-F_1&=&2 \frac{recall_{Macro}\times precision_{Macro}}{recall_{Macro}+precision_{Macro}}\text{,}\nonumber\\
precision_{Macro}&=&\frac{1}{C}\sum\limits_{i=1}^C \frac{TP_i}{(TP_i+FP_i)}\text{,}\nonumber\\ recall_{Macro}&=&\frac{1}{C}\sum\limits_{i=1}^C \frac{TP_i}{(TP_i+FN_i)}\text{,}\nonumber
\eear
\bigskip
\bear{defmacro2}
Micro-F_1&=&2 \frac{recall_{Micro}\times precision_{Micro}}{recall_{Micro}+precision_{Micro}}\text{,}\nonumber\\
precision_{Micro}&=&\frac{\sum\limits_{i=1}^C TP_i }{\sum\limits_{i=1}^C (TP_i+FP_i)}\text{,}\nonumber\\ recall_{Micro}&=&\frac{\sum\limits_{i=1}^C TP_i }{\sum\limits_{i=1}^C (TP_i+FN_i)}\text{,}\nonumber
\eear
where $C$ is the number of classes, and $TP, FP,FN$ are the numbers of true positives, false positives, false negatives, respectively. The \textit{Micro-$F_1$} measure is preferable when class sizes are imbalanced.

\begin{subsection}{Performance on the Six Clusters dataset}
	
	In this synthetic dataset, we randomly choose $10\%$ to $90\%$ of nodes for training (so as to avoid some classes missing from the training data). First, we discuss the time complexities of these embedding algorithms. The comparison of the running time of the \textit{Six Clusters} dataset is listed in Table 2. Note that our algorithm and the four baseline algorithms are all implemented in Python. The experiments are performed on an Acer Altos-T350-F2 machine with two Intel(R) Xeon(R) CPU E5-2690 v2 processors.
	
	The \textit{GME} approach finds top-\textit{k} eigenvalues by the \textit{power method} (see, e.g., the book \cite{Newman2010}), which has a time complexity of $O(|V|^2 I)$, where $I$ is the number of iterations. The time complexities of the four baseline algorithms, \textit{node2vec}, \textit{Line}, \textit{Grarep} and \textit{DeepWalk} are $O(|V|d)$, $O(|E|d)$, $O(|V|^{3})$, and $O(|V|d)$, respectively \cite{goyal2017graph}. From Table 2, it can be seen that \textit{GME} and \textit{GraRep} run faster due to the small dataset size.
	In Figure 2 (a) and (d), we show the node classification results. Since the sizes of the six classes in the \textit{Six Clusters} dataset are balanced, the two metrics \textit{Micro-$F_1$} and \textit{Macro-$F_1$} show similar results. We can see that all of the listed embedding algorithms do predict missing labels with high precision. Visualization of the \textit{GME} result of the \textit{Six Clusters} dataset using t-SNE \cite{maaten2008visualizing} is illustrated in Figure 1. Each point in Figure 1 corresponds to a node in the \textit{Six Clusters} dataset. It shows that six clusters are well separated after applying our \textit{GME} method.
	\begin{figure}[ht]
		\centering
		Figure 1. Visualization of generalized modularity embedding of \textit{Six Clusters} using t-SNE. The generalized modularity embedding is $6$-dimensional, and t-SNE reduces the dimensionality to $2$.
		\includegraphics[width=0.45\textwidth]{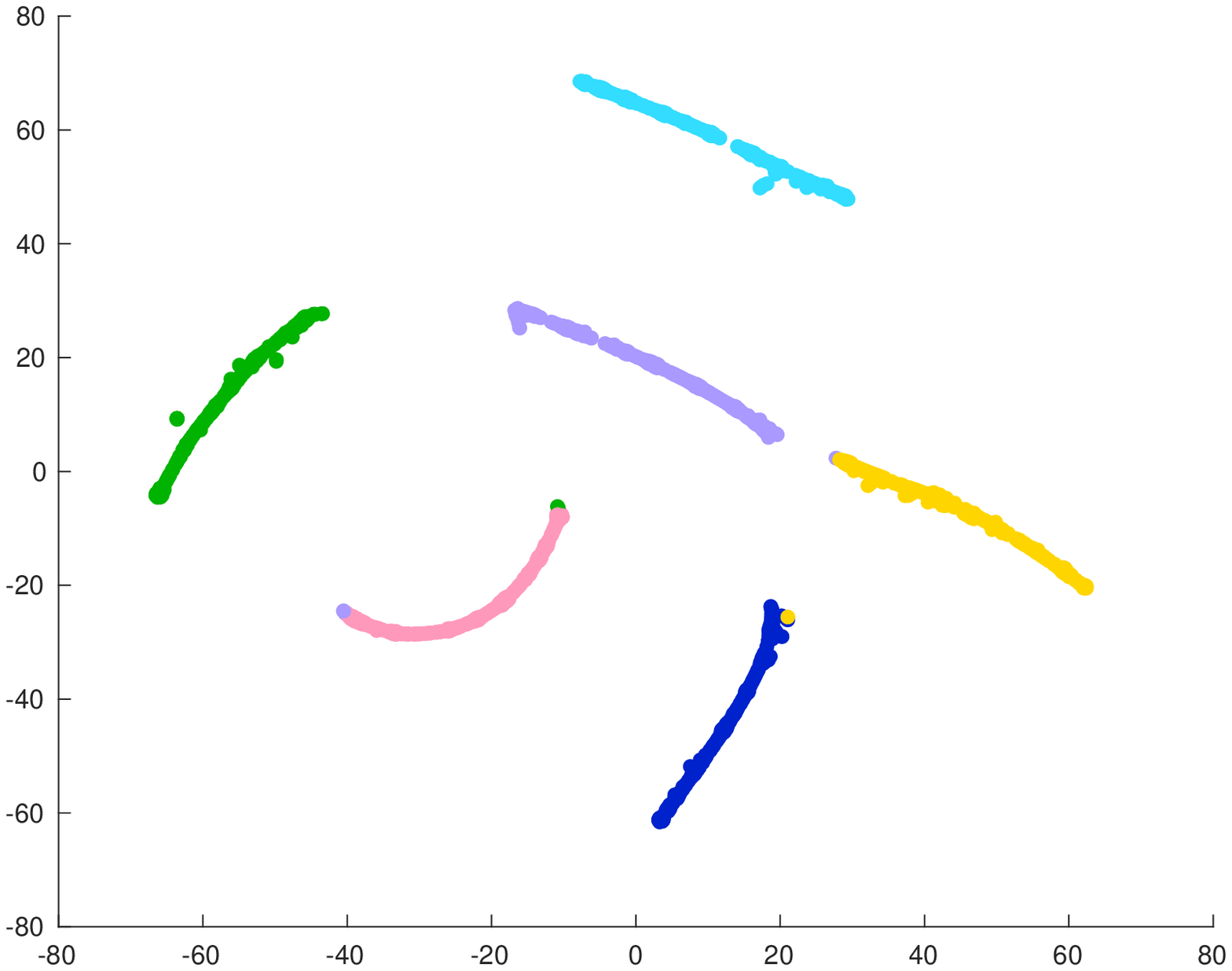}
	\end{figure}
	\begin{table*}[ht]
		\centering
		Table 2. Average running time per train-test split on the \textit{Six Clusters} dataset
		\begin{tabular}{|c||c|c|c|c|c|}
			\hline
			Algorithm & GME &node2vec& Line & GreRep	& DeepWalk \\ \hline
			Running time (s) & 22.496& 137.777& 101.965	& 17.455 & 65.285 \\ \hline
		\end{tabular}
	\end{table*}
	\begin{figure*}[ht]
		\begin{center}
			Figure 2. Performance on the \textit{Six Clusters} dataset, the \textit{Amazon} dataset and the \textit{Flickr} dataset.
			\begin{tabular}{p{0.3\textwidth}p{0.3\textwidth}p{0.3\textwidth}}
				\includegraphics[width=0.3\textwidth]{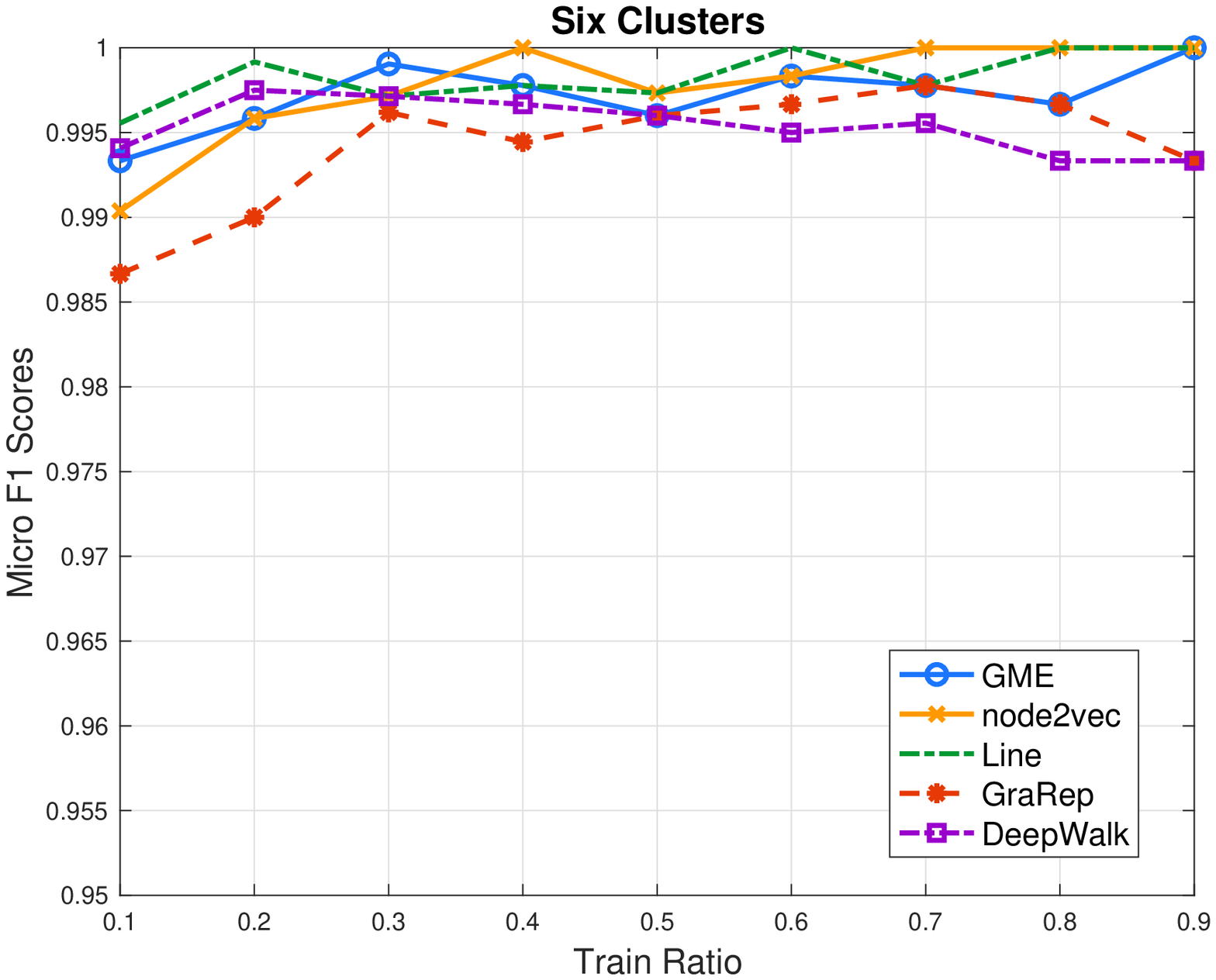} &
				\includegraphics[width=0.3\textwidth]{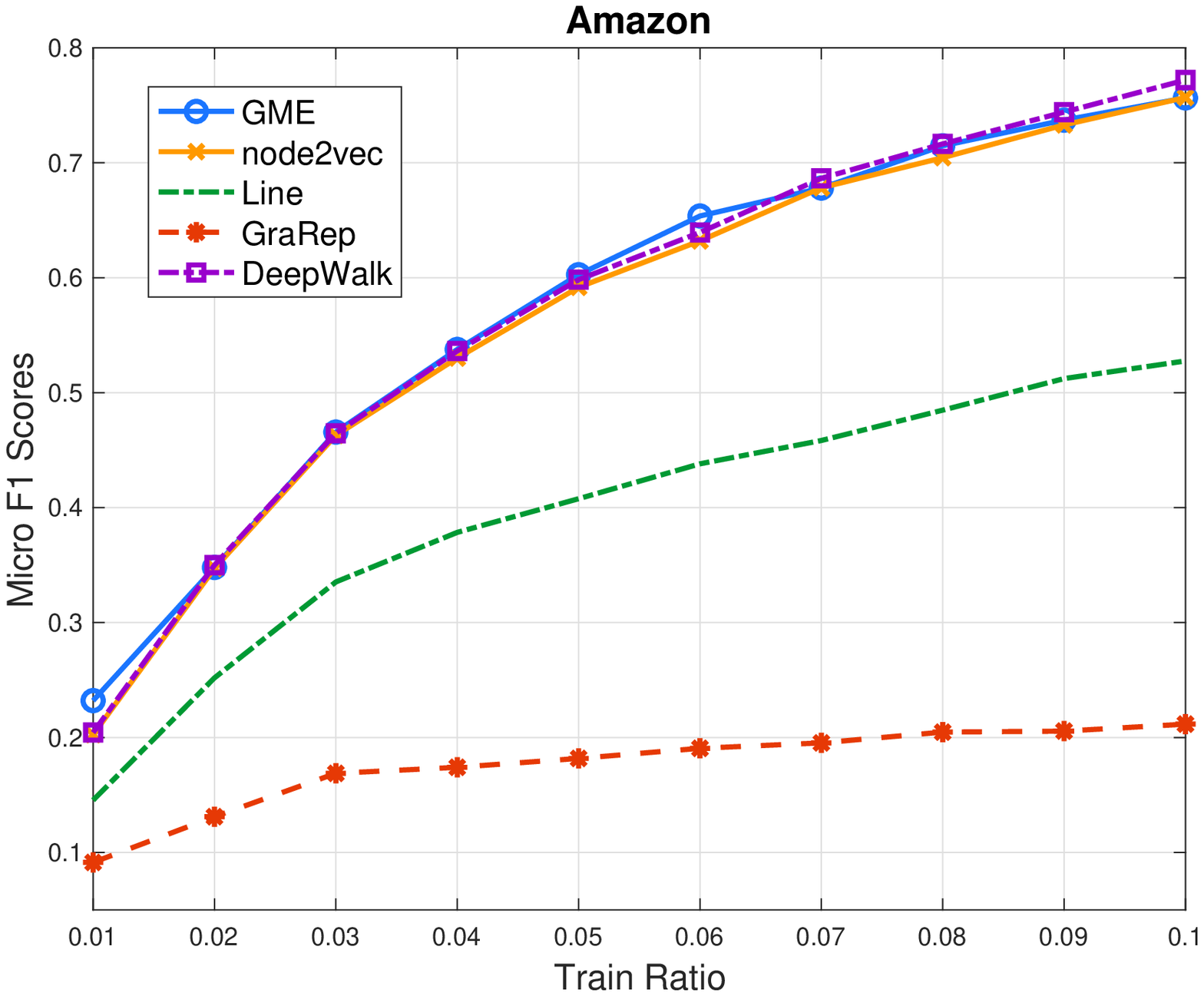} &
				\includegraphics[width=0.3\textwidth]{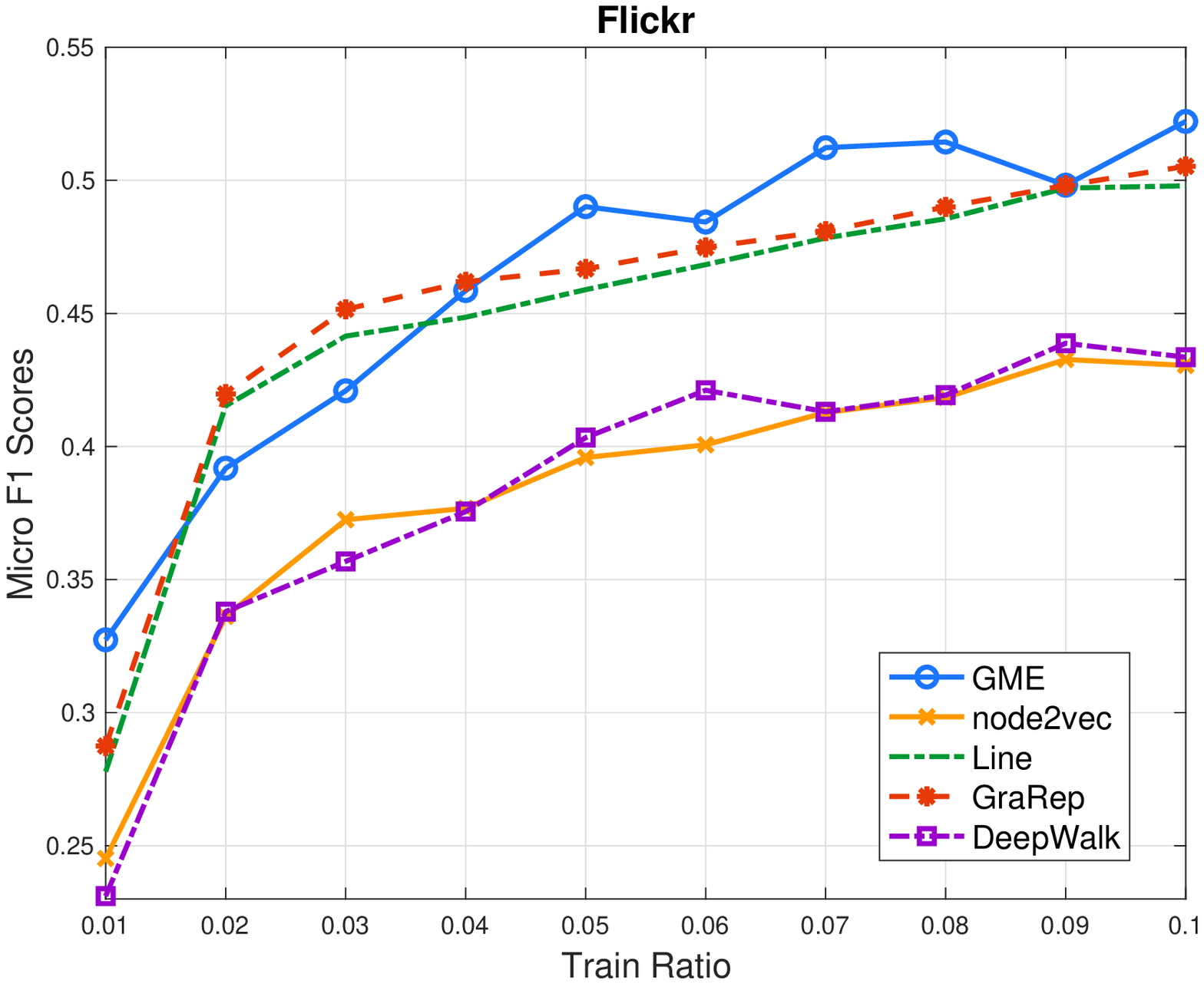}\\
				(a) Six Clusters (Micro-$F_1$) & (b) Amazon (Micro-$F_1$) & (c) Flickr (Micro-$F_1$)\\ \\
				\includegraphics[width=0.3\textwidth]{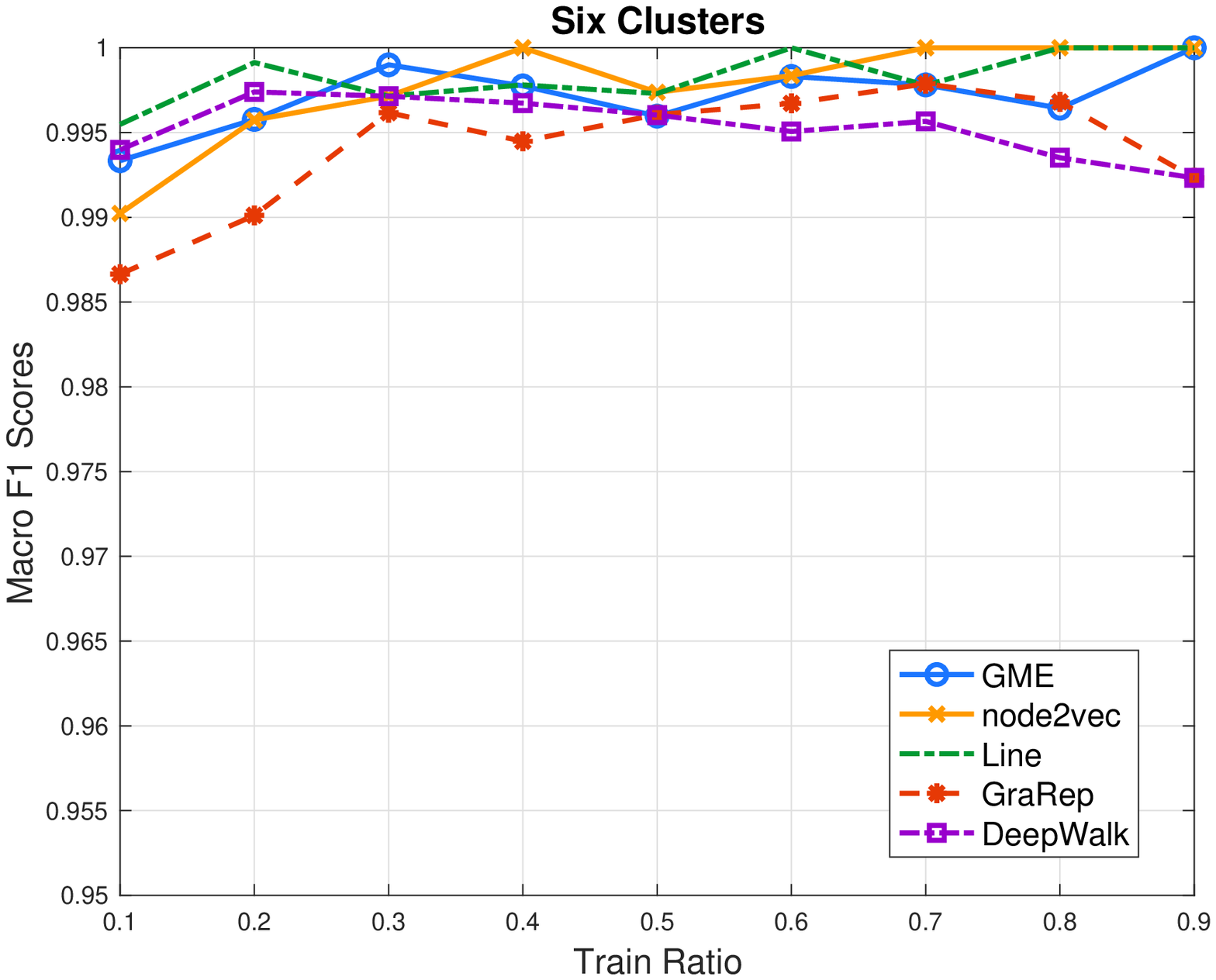}&
				\includegraphics[width=0.3\textwidth]{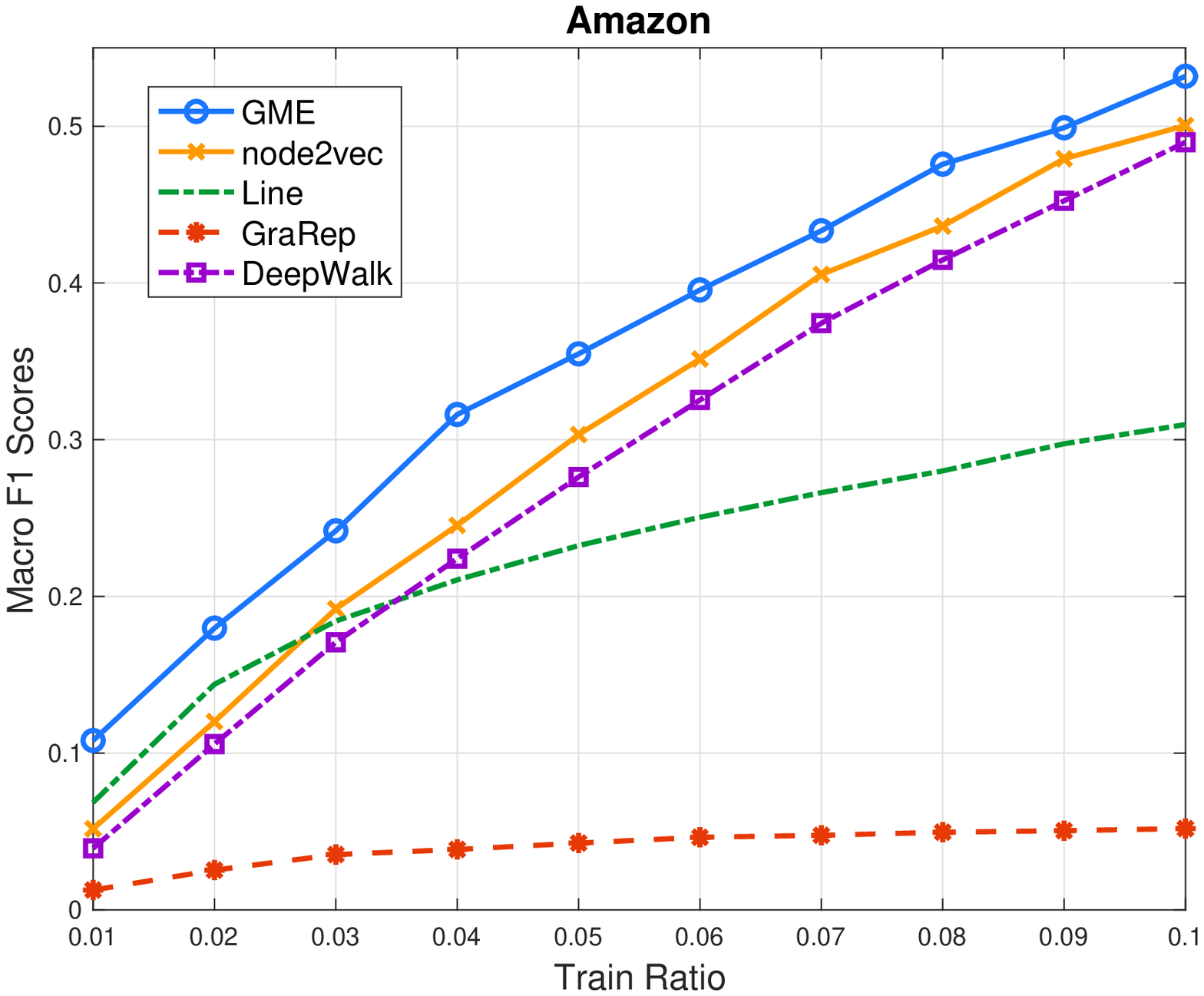} &
				\includegraphics[width=0.3\textwidth]{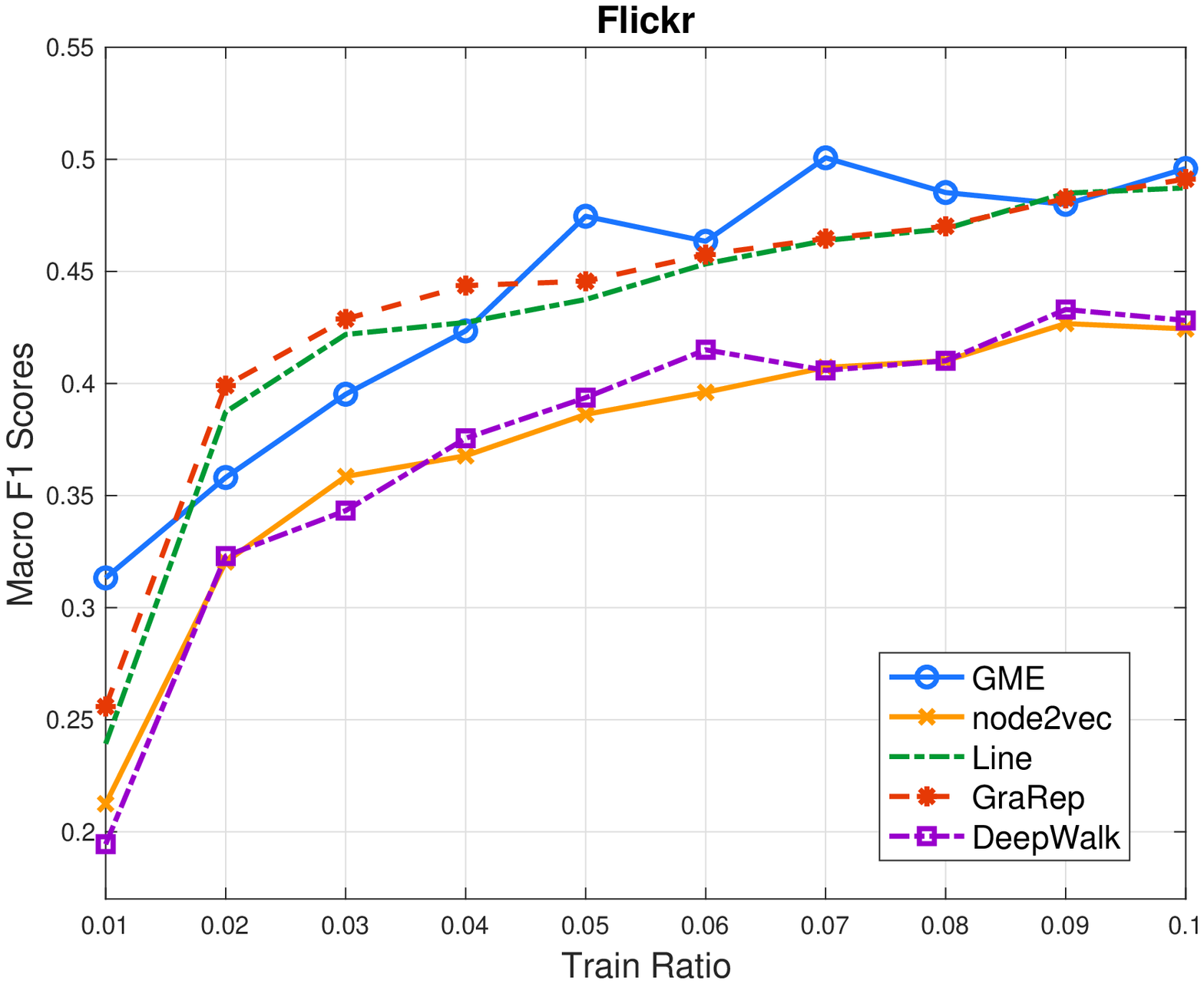}\\
				(d) Six Clusters (Macro-$F_1$) & (e) Amazon (Macro-$F_1$) & (f) Flickr (Macro-$F_1$)
			\end{tabular}
		\end{center}
	\end{figure*}
	
\end{subsection}

\begin{subsection}{Performance on the Amazon dataset}
	
	In the experiment on the \textit{Amazon} dataset, for \textit{GME}, we sample the network by random walks with different path lengths. Results are presented in Table 3. Comparing the experimental results in each train ratio, it can be seen that \textit{Micro-$F_1$} score improves as sampling path length increases. Evidently, this meets our expectations that sampling with a longer path can capture more global information.
	
	In Figure 2 (b) and (e), we show the comparison results of \textit{GME} and the other four baseline algorithms on the \textit{Amazon} dataset. \textit{Line} and \textit{GraRep} perform worse, while \textit{GME} (sampled by a random walk in our experiment) and the other two random walk based methods, \textit{DeepWalk} and \textit{node2vec}, predict missing labels relatively well. Note from Table 1 that the average degree of the \textit{Amazon} dataset is extremely sparse, as low as 5.88. The random walk approach behaves like DFS. As such, it can quickly alleviate the sparsity of the neighborhood of nodes.
	\begin{table*}[ht]\scriptsize
		\centering
		Table 3. The effect of sampling random walk path length.
		\begin{adjustbox}{width=1.\textwidth,center=\textwidth}
			\begin{tabular}{|c||c||c|c|c|c|c|c|c|c|c|c|}
				\hline
				Metric & random walk & 1\% & 2\% & 3\% & 4\% & 5\% & 6\% & 7\% & 8\% & 9\% & 10\% \\ \Xhline{2\arrayrulewidth}
				& \textit{path length 2} & 23.11 & 34.23 & 46.32 & 53.27 & 59.93 & 61.67 & 67.43 & 70.61 & 72.73 & 74.87 \\ \cline{2-12}
				Micro-$F_1$ (\%) & \textit{path length 3} & 23.21 & 34.57 & 46.56 & 53.46 & 59.13 & 62.94 & 65.86 & 71.11 & 73.40 & 75.37 \\ \cline{2-12}
				& \textit{path length 4} & \textbf{23.21} & \textbf{34.79} & \textbf{46.57} & \textbf{53.76} & \textbf{60.26} & \textbf{65.37}& \textbf{67.81} & \textbf{71.48}& \textbf{73.71} & \textbf{75.65} \\ \Xhline{2\arrayrulewidth}
				& \textit{path length 2} & 10.61 & 17.47 & 23.73 & 31.02 & 34.55 & 38.49 & 42.17 & 46.61 & 48.61 & 52.28 \\ \cline{2-12}
				Macro-$F_1$1 (\%)& \textit{path length 3} &\textbf{10.82} & 17.83 & 24.02 & 31.19 & 34.97 & 39.01 & 42.90 & 46.86 & 49.29 & 52.74 \\ \cline{2-12}
				& \textit{path length 4} & 10.81 &\textbf{17.98} & \textbf{24.19} & \textbf{31.61} & \textbf{35.48} & \textbf{39.56 } & \textbf{43.33} & \textbf{47.58} & \textbf{49.91} & \textbf{53.19} \\ \hline
			\end{tabular}
		\end{adjustbox}
	\end{table*}
\end{subsection}

\begin{subsection}{Performance on the Flickr dataset}
	
	The average degree of the \textit{Flickr} dataset is dense, which is 63.93 as summarized in Table 1. In Figure 2 (c) and (f), we show the performance results on the \textit{Flickr} dataset. Apparently, we can see that \textit{GME}, \textit{Line}, and \textit{GraRep} have better performance than \textit{DeepWalk} and \textit{node2vec}. The main reason is that random walk based approaches use random walks to enrich the neighbors of the nodes, but this may bring in noises due to the randomness of high degree nodes.
	
\end{subsection}

\bsec{Conclusion and future work}{conclusion}

In this paper, we proposed {\em Generalized Modularity Embedding} as a general framework for the network embedding problem. Our method is based on the probabilistic framework in \cite{chang2013relative, chang2017probabilistic}. By using the generalized modularity matrix as the ``similarity'' matrix in a network embedding problem, we showed that the network embedding problem can be treated as a trace maximization problem like the community detection problem. As such, it is possible to solve the network embedding problem by using the well-known community detection algorithms in the literature. On the other hand, as each sampling method leads to a sampled graph, our generalized modularity embedding approach is very general and flexible. In particular, we showed that the Laplacian eigenmaps is a special case of our generalized modularity embedding. Also, we showed that dimensionality reduction can be done by using a particular sampled graph. Various experiments were conducted on a synthetic dataset and two real datasets to illustrate the effectiveness of our approach.

For future works, we plan to extend the general modularity embedding framework to attributed networks where there are edge attributes and node attributes representing the “features” of edges and nodes \cite{chang2018e}. Also, we would like to study possible feedback mechanisms that can automatically learn from the dataset to select an appropriate sampling method (the view point) for a specific task.

\newpage

\section*{Appendix A: proof of \rthe{embed}}

(i) Note that the constraints in \req{embed1122} and \req{embed1133} are equivalent to that the $K$ columns of $H$ are orthonormal vectors, i.e., $H^T H={\bf I}_K$.

Since $\sum_{u=1}^n \qq(u,w)=\sum_{w=1}^n \qq_{u,w}=0$, it follows that
\bear{embed2222}
&& \sum_{u=1}^n \sum_{w=1}^n \qq(u,w) ||h_u-h_w||^2 \nonumber \\
&&=-2\sum_{u=1}^n \sum_{w=1}^n \qq(u,w) (h_u^T \cdot h_w) \label{eq:embed2233} \\
&&=-2 \sum_{u=1}^n \sum_{w=1}^n \sum_{k=1}^K \qq(u,w) h_{u,k}h_{w,k} \nonumber \\
&&=-2 \mbox{tr}(H^T \Q H),
\eear
where $H$ is the $n \times K$ matrix with its $u^{th}$ row being $h_u$. Thus, the minimization problem in \req{embed1111} is the same as the maximization problem in \req{opti1111tr}.

(ii) Note that
$$||\Q- H H^T||_2^2 =\sum_{u=1}^n \sum_{w=1}^n (q(u,w)-h_u^T h_w)^2,$$
where $h_u=(h_{u,1}, h_{u,2}, \ldots, h_{u,K})^T$ is the transpose of the $u^{th}$ row vector of $H$. Also, the constraint $H^T H={\bf I}_K$ is equivalent to
\beq{decom0000}
\sum_{u=1}^n h_{u,k}^2 =1, \quad k=1, 2, \ldots, K,
\eeq
and
\beq{decom0011}
\sum_{u=1}^n h_{u,k_1} h_{u,k_2}=0, \quad k_1 \ne k_2.
\eeq
It then follows that
\bear{decom1111}
||\Q- H H^T||_2^2 &=&\sum_{u=1}^n \sum_{w=1}^n (q(u,w)-h_u^T h_w)^2 \nonumber \\
&=&\sum_{u=1}^n \sum_{w=1}^n q(u,w)^2 \nonumber \\
&&- 2 \sum_{u=1}^n \sum_{w=1}^n q(u,w)(h_u^T h_w) \nonumber \\
&&+\sum_{u=1}^n \sum_{w=1}^n (h_u^T h_w)^2.
\eear
Note that the first term in \req{decom1111} is a constant. Since $h_u^T h_w=\sum_{k=1}^K h_{u,k}h_{w,k}$, we have for the third term in \req{decom1111} that
\bearn
&&\sum_{u=1}^n \sum_{w=1}^n (h_u^T h_w)^2 \\
&=&\sum_{u=1}^n \sum_{w=1}^n (\sum_{k=1}^K h_{u,k}h_{w,k})^2 \\
&=&\sum_{u=1}^n \sum_{w=1}^n \sum_{k_1=1}^K \sum_{k_2=1}^K h_{u,k_1}h_{w,k_1}h_{u,k_2}h_{w,{k_2}}\\
&=&\sum_{u=1}^n \sum_{w=1}^n \sum_{k_1=1}^K h_{u,k_1}^2 h_{w,k_1}^2\nonumber\\
&+&\sum_{u=1}^n \sum_{w=1}^n \sum_{k_1=1}^K \sum_{k_2 \ne k_1} h_{u,k_1}h_{w,k_1}h_{u,k_2}h_{w,{k_2}} \\
\eearn
From \req{decom0000}, we have
$$\sum_{u=1}^n \sum_{w=1}^n \sum_{k_1=1}^K h_{u,k_1}^2 h_{w,k_1}^2=K.$$
Also, we have from \req{decom0011} that
$$\sum_{u=1}^n \sum_{w=1}^n \sum_{k_1=1}^K \sum_{k_2 \ne k_1} h_{u,k_1}h_{w,k_1}h_{u,k_2}h_{w,{k_2}}=0.$$
Thus, $\sum_{u=1}^n \sum_{w=1}^n (h_u^T h_w)^2$ is also a constant. It then follows from \req{decom1111} and \req{embed2233} that the minimization problem in \req{embed1111} is the same as the minimization problem in \req{opti1111decom}.

\section*{Appendix B: proof of \rthe{ObjIncreasing}}

For the proof of \rthe{ObjIncreasing}, we need the following properties in \rlem{StrictlyIncreasing} to show that the objective function is increasing after each update and thus converges to a local optimum in \rthe{ObjIncreasing}.

\begin{lemma}\label{lem:StrictlyIncreasing}(\cite{ProfSoftmax}, Lemma 8.1.5)
	Suppose $X$ is a random variable with the probability mass function $P(X=k)=p_k$, $k=1, 2, \ldots, K$, and that $g(X)$ is not a constant (w.p.1) for some $g: \Re\mapsto\Re$. Let
	\begin{equation}
		\Lambda(\theta)=\log\big(\mathbf{E}[e^{\theta g(X)}]\big).
	\end{equation}
	Then for all $\theta > 0$,
	$$\frac{\mathbf{E}[g(X)e^{\theta g(X)}]}{\mathbf{E}[e^{\theta g(X)}]} > \mathbf{E}[g(X)].$$
\end{lemma}

Now we prove \rthe{ObjIncreasing}. Suppose that $u_0$ is the point updated in Step (7) of Algorithm \ref{alg:Softmax}.
After the update, the objective value in \req{objc1111} can be written as follows:
\bear{objc2222}
\sum\limits_{k=1}^K\bigg(&\sum\limits_{u\neq u_0}\sum\limits_{w\neq u_0}q(u,w)h_{u,k}h_{w,k} \nonumber\\
&+ \sum\limits_{w\neq u_0}q(u_0,w) \tilde{h}_{u_0,k}h_{w,k} \nonumber \\
&+ \sum\limits_{u\neq u_0}q(u, u_0)h_{u,k}\tilde{h}_{u_0,k} \nonumber \\
&+ q(u_0,u_0)\tilde{h}_{u_0,k}\tilde{h}_{u_0,k}\bigg).
\eear
Since $q(u,w)=q(w,u)$ (symmetric) and $q(u,u)=0$ for all $u$, we have from \req{objc2222} that the difference, denoted by $\Delta_{Obj}$, between the objective value {\em after} the update and that {\em before} the update can be computed as follows:
\begin{equation}
	\begin{aligned}
		\Delta_{Obj} &= \sum\limits_{k=1}^K 2\cdot\bigg[\sum\limits_{w\neq u_0}q(u_0,w)\tilde{h}_{u_0,k}h_{w,k}\\
		&-\sum\limits_{w\neq u_0}q(u_0,w) h_{u_0,k}h_{w,k}\bigg] \\
		&= 2\cdot\sum\limits_{k=1}^K\bigg[\tilde{h}_{u_0,k}\bigg(\sum\limits_{w\neq u_0}q(u_0,w)h_{w,k}\bigg)\\
		&- h_{u_0,k}\bigg(\sum\limits_{w\neq u_0}q(u_0, w)h_{w,k}\bigg)\bigg] \\
		&= 2\cdot\sum\limits_{k=1}^K\bigg[\tilde{h}_{u_0,k}z_{u_0,k} - h_{u_0,k}z_{u_0,k}\bigg].
	\end{aligned}
\end{equation}
Now view $z_{u_0,k}$ as $g(k)$, $h_{u_0,k}$ as $P(X=k)$ in \rlem{StrictlyIncreasing}.

Then
\beq{objc3333}
\tilde{h}_{u_0,k} = \frac{e^{\theta z_{u_0,k}}h_{u_0,k}}{\sum\limits_{\ell=1}^K e^{\theta z_{u_0,\ell}}h_{u_0,\ell}} = \frac{e^{\theta g(k)}h_{u_0,k}}{\mathbf{E}[e^{\theta g(X)}]}.
\eeq
Note from \req{objc3333} that $\Delta_{Obj}$ can also be written as follows:
\begin{equation}
	\begin{aligned}
		\Delta_{Obj} &= 2\cdot\bigg[\sum\limits_{k=1}^K z_{u_0,k}\frac{e^{\theta z_{u_0,k}}h_{u_0,k}}{\sum\limits_{\ell=1}^K e^{\theta z_{u_0,\ell}}h_{u_0,\ell}} - \sum\limits_{k=1}^K h_{u_0,k}z_{u_0,k}\bigg] \\
		&= 2\bigg[\frac{\mathbf{E}[g(X)e^{\theta g(X)}]}{\mathbf{E}[e^{\theta g(X)}]} - \mathbf{E}[g(X)]\bigg].
	\end{aligned}
\end{equation}
From Lemma \ref{lem:StrictlyIncreasing}, we conclude that $\Delta_{Obj}>0$ for all $\theta>0$. Since the objective values in \req{objc1111} are bounded, the the objective values converge monotonically to a finite constant.


\begin{thebibliography}{99}
	\providecommand{\url}[1]{#1}
	\csname url@samestyle\endcsname
	\providecommand{\newblock}{\relax}
	\providecommand{\bibinfo}[2]{#2}
	\providecommand{\BIBentrySTDinterwordspacing}{\spaceskip=0pt\relax}
	\providecommand{\BIBentryALTinterwordstretchfactor}{4}
	\providecommand{\BIBentryALTinterwordspacing}{\spaceskip=\fontdimen2\font plus
		\BIBentryALTinterwordstretchfactor\fontdimen3\font minus
		\fontdimen4\font\relax}
	\providecommand{\BIBforeignlanguage}[2]{{
			\expandafter\ifx\csname l@#1\endcsname\relax
			\typeout{** WARNING: IEEEtran.bst: No hyphenation pattern has been}
			\typeout{** loaded for the language `#1'. Using the pattern for}
			\typeout{** the default language instead.}
			\else
			\language=\csname l@#1\endcsname
			\fi
			#2}}
	\providecommand{\BIBdecl}{\relax}
	\BIBdecl
	
	\bibitem{Newman2010}
	M.~Newman, \emph{Networks: an introduction}.\hskip 1em plus 0.5em minus 0.4em\relax OUP Oxford, 2009.
	
	\bibitem{goodfellow2016deep}
	I.~Goodfellow, Y.~Bengio, A.~Courville, and Y.~Bengio, \emph{Deep learning}.\hskip 1em plus 0.5em minus 0.4em\relax MIT press Cambridge, 2016, vol.~1.
	
	\bibitem{perozzi2014deepwalk}
	B.~Perozzi, R.~Al-Rfou, and S.~Skiena, ``Deepwalk: Online learning of social representations,'' in \emph{Proceedings of the 20th ACM SIGKDD international conference on Knowledge discovery and data mining}.\hskip 1em plus 0.5em minus 0.4em\relax ACM, 2014, pp. 701--710.
	
	\bibitem{tang2015line}
	J.~Tang, M.~Qu, M.~Wang, M.~Zhang, J.~Yan, and Q.~Mei, ``Line: Large-scale information network embedding,'' in \emph{Proceedings of the 24th International Conference on World Wide Web}.\hskip 1em plus 0.5em minus 0.4em\relax International World Wide Web Conferences Steering Committee, 2015, pp. 1067--1077.
	
	\bibitem{cao2015grarep}
	S.~Cao, W.~Lu, and Q.~Xu, ``Grarep: Learning graph representations with global structural information,'' in \emph{Proceedings of the 24th ACM International on Conference on Information and Knowledge Management}.\hskip 1em plus 0.5em minus 0.4em\relax ACM, 2015, pp. 891--900.
	
	\bibitem{grover2016node2vec}
	A.~Grover and J.~Leskovec, ``node2vec: Scalable feature learning for networks,'' in \emph{Proceedings of the 22nd ACM SIGKDD international conference on Knowledge discovery and data mining}.\hskip 1em plus 0.5em minus 0.4em\relax ACM, 2016, pp. 855--864.
	
	\bibitem{wang2016structural}
	D.~Wang, P.~Cui, and W.~Zhu, ``Structural deep network embedding,'' in \emph{Proceedings of the 22nd ACM SIGKDD international conference on Knowledge discovery and data mining}.\hskip 1em plus 0.5em minus 0.4em\relax ACM, 2016, pp. 1225--1234.
	
	\bibitem{ou2016asymmetric}
	M.~Ou, P.~Cui, J.~Pei, Z.~Zhang, and W.~Zhu, ``Asymmetric transitivity preserving graph embedding,'' in \emph{Proceedings of the 22nd ACM SIGKDD international conference on Knowledge discovery and data mining}.\hskip 1em plus 0.5em minus 0.4em\relax ACM, 2016, pp. 1105--1114.
	
	\bibitem{goyal2017graph}
	P.~Goyal and E.~Ferrara, ``Graph embedding techniques, applications, and performance: A survey,'' \emph{arXiv preprint arXiv:1705.02801}, 2017.
	
	\bibitem{wang2017community}
	X.~Wang, P.~Cui, J.~Wang, J.~Pei, W.~Zhu, and S.~Yang, ``Community preserving network embedding.'' in \emph{AAAI}, 2017, pp. 203--209.
	
	\bibitem{lai2017preserving}
	Y.-A. Lai, C.-C. Hsu, W.~H. Chen, M.-Y. Yeh, and S.-D. Lin, ``Preserving proximity and global ranking for node embedding,'' in \emph{Advances in Neural Information Processing Systems}, 2017, pp. 5261--5270.
	
	\bibitem{tenenbaum2000global}
	J.~B. Tenenbaum, V.~De~Silva, and J.~C. Langford, ``A global geometric framework for nonlinear dimensionality reduction,'' \emph{Science}, vol. 290, no. 5500, pp. 2319--2323, 2000.
	
	\bibitem{belkin2002laplacian}
	M.~Belkin and P.~Niyogi, ``Laplacian eigenmaps and spectral techniques for embedding and clustering,'' in \emph{Advances in Neural Information Processing Systems}, 2002, pp. 585--591.
	
	\bibitem{li2013modularity}
	W.~Li, ``Modularity embedding,'' in \emph{International Conference on Neural Information Processing}.\hskip 1em plus 0.5em minus 0.4em\relax Springer, 2013, pp. 92--99.
	
	\bibitem{chang2013relative}
	C.-S. Chang, C.-J. Chang, W.-T. Hsieh, D.-S. Lee, L.-H. Liou, and W.~Liao, ``Relative centrality and local community detection,'' \emph{Network Science}, vol.~3, no.~4, pp. 445--479, 2015.
	
	\bibitem{chang2017probabilistic}
	C.-S. Chang, D.-S. Lee, L.-H. Liou, S.-M. Lu, and M.-H. Wu, ``A probabilistic framework for structural analysis and community detection in directed networks,'' \emph{IEEE/ACM Transactions on Networking}, vol.~26, no.~1, pp. 31--46, 2018.
	
	\bibitem{Newman04}
	M.~E. Newman, ``Fast algorithm for detecting community structure in networks,'' \emph{Physical Review E}, vol.~69, no.~6, p. 066133, 2004.
	
	\bibitem{Lambiotte2010}
	R.~Lambiotte, ``Multi-scale modularity in complex networks,'' in \emph{Modeling and Optimization in Mobile, Ad Hoc and Wireless Networks (WiOpt), 2010 Proceedings of the 8th International Symposium on}.\hskip 1em plus 0.5em minus 0.4em\relax IEEE, 2010, pp. 546--553.
	
	\bibitem{Delvenne2010}
	J.-C. Delvenne, S.~N. Yaliraki, and M.~Barahona, ``Stability of graph communities across time scales,'' \emph{Proceedings of the National Academy of Sciences}, vol. 107, no.~29, pp. 12\,755--12\,760, 2010.
	
	\bibitem{FB07}
	S.~Fortunato and M.~Barthelemy, ``Resolution limit in community detection,'' \emph{Proceedings of the National Academy of Sciences}, vol. 104, no.~1, pp. 36--41, 2007.
	
	\bibitem{blondel2008fast}
	V.~D. Blondel, J.-L. Guillaume, R.~Lambiotte, and E.~Lefebvre, ``Fast unfolding of communities in large networks,'' \emph{Journal of Statistical Mechanics: Theory and Experiment}, vol. 2008, no.~10, p. P10008, 2008.
	
	\bibitem{shi2000normalized}
	J.~Shi and J.~Malik, ``Normalized cuts and image segmentation,'' \emph{Pattern Analysis and Machine Intelligence, IEEE Transactions on}, vol.~22, no.~8, pp. 888--905, 2000.
	
	\bibitem{yu2003multiclass}
	S.~X. Yu and J.~Shi, ``Multiclass spectral clustering,'' in \emph{Computer Vision, 2003. Proceedings. Ninth IEEE International Conference on}.\hskip 1em plus 0.5em minus 0.4em\relax IEEE, 2003, pp. 313--319.
	
	\bibitem{Dhillon04b}
	I.~Dhillon, Y.~Guan, and B.~Kulis, \emph{A unified view of kernel k-means, spectral clustering and graph cuts}.\hskip 1em plus 0.5em minus 0.4em\relax Computer Science Department, University of Texas at Austin, 2004.
	
	\bibitem{chang2016mathematical}
	C.-S. Chang, W.~Liao, Y.-S. Chen, and L.-H. Liou, ``A mathematical theory for clustering in metric spaces,'' \emph{IEEE Transactions on Network Science and Engineering}, vol.~3, no.~1, pp. 2--16, 2016.
	
	\bibitem{von2007tutorial}
	U.~Von~Luxburg, ``A tutorial on spectral clustering,'' \emph{Statistics and Computing}, vol.~17, no.~4, pp. 395--416, 2007.
	
	\bibitem{chang2017k}
	C.-S. Chang, C.-T. Chang, D.-S. Lee, and L.-H. Liou, ``K-sets+: a linear-time clustering algorithm for data points with a sparse similarity measure,'' in \emph{Proceedings of the 3rd IEEE Int'l Conf. on Cloud and Big Data Computing (CBDCom)}.\hskip 1em plus 0.5em minus 0.4em\relax IEEE, 2017.
	
	\bibitem{chang2018e}
	C.-H. Chang, C.-S. Chang, C.-T. Chang, D.-S. Lee, and P.-E. Lu, ``Exponentially twisted sampling for centrality analysis and community detection in attributed networks,'' in \emph{Proceedings of IEEE ICC 2018}, May 2018.
	
	\bibitem{spielman2007spectral}
	D.~A. Spielman, ``Spectral graph theory and its applications,'' in \emph{Foundations of Computer Science, 2007. FOCS'07. 48th Annual IEEE Symposium on}.\hskip 1em plus 0.5em minus 0.4em\relax IEEE, 2007, pp. 29--38.
	
	\bibitem{SoftmaxFunc}
	C.~M. Bishop, ``Pattern recognition,'' \emph{Machine Learning}, vol. 128, pp. 1--58, 2006.
	
	\bibitem{flickrdataset}
	X.~Huang, J.~Li, and X.~Hu, ``Accelerated attributed network embedding,'' in \emph{SIAM International Conference on Data Mining}, 2017, pp. 633--641.
	
	\bibitem{yang2015defining}
	J.~Yang and J.~Leskovec, ``Defining and evaluating network communities based on ground-truth,'' \emph{Knowledge and Information Systems}, vol.~42, no.~1, pp. 181--213, 2015.
	
	\bibitem{maaten2008visualizing}
	L.~v.~d. Maaten and G.~Hinton, ``Visualizing data using t-sne,'' \emph{Journal of machine learning research}, vol.~9, no. Nov, pp. 2579--2605, 2008.
	
	\bibitem{ProfSoftmax}
	C.-S. Chang, \emph{Performance guarantees in communication networks}.\hskip 1em plus 0.5em minus 0.4em\relax Springer Science \& Business Media, 2012.
\end{thebibliography}
\end{document}